\documentclass[aps,prl,preprint,groupedaddress,lineno]{revtex4-1}
\usepackage{mathrsfs}
\usepackage{graphicx}% Include figure files
\usepackage{dcolumn}% Align table columns on decimal pointTyr
\usepackage{bm}% bold math
\usepackage{lineno}
\usepackage{color}

\begin{document}

\title{Uncovering the molecular mechanism for dual effect of ATP on phase separation in FUS solution}

\author{Chun-Lai Ren$^{1,\dag,*}$, Yue Shan$^{1,\dag}$, Pengfei Zhang$^2$, Hong-Ming Ding$^{3,*}$, and Yu-Qiang Ma$^{1,*}$}

\affiliation{$^1$ National Laboratory of Solid State Microstructures and Department of Physics, Collaborative Innovation Center of Advanced Microstructures, Nanjing University, Nanjing 210093, China \\
                $^2$ State Key Laboratory for Modification of Chemical Fibers and Polymer Materials, Center for Advanced Low-Dimension Materials, College of Material Science and Engineering, Donghua University, Shanghai 201620, China \\
                $^3$ Center for Soft Condensed Matter Physics and Interdisciplinary Research, School of Physical Science and Technology, Soochow University, Suzhou 215006, China. \\
                $^\dag$ These authors contributed equally to this work.  \\
                $^*$ To whom correspondence should be addressed. E-mail: chunlair@nju.edu.cn (C. R.); dinghm@suda.edu.cn (H. D.); myqiang@nju.edu.cn (Y. M.)}

%\date{\today}

\begin{abstract}
Recent studies reported that adenosine triphosphate (ATP) could inhibit as well as enhance the phase separation in prion-like proteins. The molecular mechanism underlying such a puzzling phenomenon remains elusive. Here, taking the fused in sarcoma (FUS) solution as an example, we comprehensively reveal the underlying mechanism by which ATP regulates phase separation by combining the semiempirical quantum mechanical  method, mean-field theory, and molecular simulation. At the microscopic level, ATP acts as a bivalent or trivalent binder; at the macroscopic level, the reentrant phase separation indeed occurs in dilute FUS solutions, resulting from the ATP-concentration--dependent binding ability under different conditions. Importantly, the ATP concentration for dissolving the protein condensates is about 10 $mM$, agreeing with experimental results. Furthermore, from a dynamic point of view, the effect of ATP on phase separation is also non-monotonic. This work provides a clear physical description of the microscopic interaction and macroscopic phase diagram of the ATP-modulated phase separation.

\end{abstract}

\maketitle

\section{Teaser}
The multivalence of ATP induces a concentration-dependent binding ability and contributes to its dual role in phase separation.

\section{Introduction}
Recent discoveries have shown that phase separation plays an important role in many intracellular biological processes \cite{nature,brang1,brang2,brang3,brang4,brang5}, e.g., the formation of membrane-less compartments like stress granules in the cytoplasm \cite{sg}, Cajal bodies \cite{cb1}, and nucleoli \cite{nu} in the nucleus. Phase separation results in the compartmentalization of biomolecules into concentrated liquid droplets (with a typical size in micrometers) via weak specific interactions at the molecular level \cite{mi2,IDR1,IDR2}.  Although their organizations are typically highly dynamic and accompanied by the rapid exchange of components with their surroundings, their overall size and shape are usually stable for minutes or hours \cite{stab1,stab2}.

The development of both experiment and theory \cite{PRL1,PNAS1,CB,berry,brang7,pappu1,pappu2,wingreen1} has revealed the possible mechanisms of phase separation. One central principle is multivalence, which can drive the formation of condensate via phase separation coupled to percolation \cite{PSCP1,PSCP2,PSCP3,PSCP4,PSCP5}. Specifically, multivalent interaction is facilitated by the intrinsically disordered regions (IDRs) in the proteins \cite{FUS1,FUS2,FUS3}.  As one specific class of IDRs, the prion-like domain (PLD) is frequently found in RNA-binding proteins, e.g., the fused in sarcoma (FUS) protein. It was found that the molecular driving forces for phase separation in FUS were cooperatively weak interactions among tyrosine (Tyr) residues from the PLD and arginine (Arg) residues from the RNA binding domain (RBD) \cite{FUS4}; the valence of the multivalent interactions was a key factor determining the saturation concentration in FUS solutions.

Adenosine triphosphate (ATP) is known as an energy fuel for biological reactions typically at the concentration of micromoles ($\mu M$) \cite{SG1,SG2,SG3}. In cells, surprisingly, the ATP concentration is in the range of 2--12 millimoles ($mM$), which is much higher than the amount required for the energy fuel. Previous studies showed that the high concentration of ATP was related to the dissolution of protein condensates \cite{science1}, in which ATP was explained as a biological hydrotrope \cite{hydrotrope2,hydrotrope3,hydrotrope4}. In chemistry, hydrotropes are small-molecule amphiphiles that contain distinct hydrophilic and hydrophobic parts (the aromatic ring/rings in most cases) and can help solubilize the hydrophobic substances in aqueous systems \cite{hydrotrope1}. Likewise, ATP consists of hydrophobic aromatic pyrimidine base ring and hydrophilic triphosphate moiety, fitting well into the concept of a hydrotrope.

However, later works found that like RNA \cite{RNA}, ssDNA \cite{plos} and other molecules \cite{sm}, ATP had two-stage effects on phase separation of prion-like proteins: an enhancement of phase separation at low concentrations and an inhibition of phase separation at high concentrations \cite{plos,kang,song,dual}. The mechanism of these dual effects of ATP is still largely unknown. It is worth noting that such reentrant phase separations are quite general in biomacromolecules, which were observed in many multi-component bio-systems \cite{reentrant0,epyc2,spop,reentrant1,reentrant3}.  Importantly, there is emerging evidence that abnormal protein aggregation is associated with many human diseases, including cancer, neurodegeneration, and infectious disease \cite{stab2,dise}. Therefore, it is of fundamental and physiological significance to uncover the influence of ATP on protein condensation or solubilization.

In this work, we combine the semiempirical quantum mechanical (SQM) method, mean-field theory, and coarse-grained molecular dynamics (CGMD) simulation to reveal the mechanism underlying the dual effect of ATP on phase separation. We first use the SQM method to uncover the Tyr--Arg and ATP--Arg bindings at the atomic level, which are taken as the input for the mean-field theory. We then develop a mean-field theory by explicitly taking molecular interactions into account based on a stickers-and-spacers model, and provide phase diagrams with full binodals at the macroscopic level, which can be directly compared with the experimental phenomena. Furthermore, we employ the CGMD simulation to investigate the phase separation from the dynamic perspective. These three methods complement each other and can build a direct relationship between the microscopic driving force and macroscopic phase separation, offering a better understanding of ATP-modulated  phase separation.

\section{Results}

\textbf{The relationship between Tyr--Arg binding and phase separation in FUS solution.} Multivalent interactions among Tyr and Arg residues are considered as driving forces for the phase separation in FUS solution \cite{FUS4}. However, the direct relationship between the multivalent interactions and phase diagrams is still lacking. To achieve a quantitative description, we first employed the semi-empirical quantum mechanical (SQM) method to calculate the binding free energy of the Tyr--Arg interaction (more details on the SQM method are given in the Materials and Methods). Fig. 1A--B shows two typical binding modes of the Tyr--Arg complex, where the ending charged group in Arg may be either perpendicular or parallel to the benzene plane in Tyr. Notably, the binding free energy is about 2.0 $k_BT$ in both cases, indicating that the Tyr--Arg binding is a weak interaction. This raises the question of whether  Tyr--Arg binding could be sufficient to cause the phase separation in FUS solutions.

To answer this question, we developed a mean-field theory for FUS solutions based on a stickers-and-spacers model. This model was first proposed by Semenov and Rubinstein to treat the gelation and phase separation of thermo-responsive associative polymer solutions  \cite{the1}, following the pioneering works by Tanaka for similar systems \cite{FTanaka1,FTanaka2}. Recently, this model has been successfully used to reflect the phase separation in multivalent proteins \cite{pappu1,pappu2,pappu3}. It should be mentioned that two order parameters, i.e., the volume fraction of protein and the fraction of Tyr--Arg bonds, self-consistently appear in the free energy, which is widely used to reflect the concentration and structure in the molecular liquid \cite{the1,Tanaka1}. To better compare the theoretical and experimental results, we convert the calculated volume fraction into the concentration (see details of the Theoretical modeling in the Materials and Methods section) in the following discussion. Fig. 1C shows the phase diagram (i.e., the binodal curve) of phase separation in the FUS solution, depending on the FUS concentration ($\rho$) and the Tyr--Arg binding free energy ($\epsilon_1$). The details of the calculation for the phase diagram are given in the Supplementary Materials. When $\epsilon_1$ is lower than $-1.4 k_BT$, the phase separation could be observed in the FUS solution, meaning that the Tyr--Arg binding is strong enough to drive phase separation due to the cooperation of multiple Tyr--Arg bindings. At $\epsilon_1=-2.0 k_BT$, the concentration of the condensed phase is two orders of magnitude higher than that of the dilute phase, which is consistent with the characteristics of protein condensates \cite{IDR2}. Because the binding free energy of Tyr--Arg is weak, there may exist a reversible breaking and formation of the Tyr--Arg binding under the thermal fluctuation. The extent of Tyr--Arg binding can be reflected by the fraction of Tyr--Arg bonds $p$, namely the ratio of the number of Tyr--Arg bonds to the number of Tyr residues in the solution. As shown in Fig. 1D, the value of $p$ in the dense phase is larger than that in the diluted phase, indicating that the protein condensates are denser liquids with network structures in which Tyr--Arg bonds act as cross-linkers (Fig. 1E). Collectively, the above results demonstrate that different FUS concentrations, as well as distinct microscopic structures, occur in the two coexisting phases.

\textbf{SQM methods on the microscopic interactions between ATP and Arg.} We then investigated the effect of ATP on the phase separation in FUS solutions. With the addition of ATP, there may exist an interaction between ATP and FUS, in which the ATP--Arg binding is believed to play an important role \cite{plos,kang,song,dual}. However, the binding modes, possible binding numbers of Arg residues (to ATP), and binding strength are still lacking. To reveal the detailed microscopic mechanism of their interactions, the SQM method was again employed. One study \cite{plos} assumed that ATP may act as a bivalent binder due to its amphiphilic nature, i.e., two Arg residues could bind to the triphosphate part and the adenine part, respectively. As shown in Fig. 2A--B, the Arg can indeed bind to either the triphosphate part or the adenine part of ATP. Nevertheless, the binding strength is quite different in the two cases. Since the Arg carries one net positive charge and the triphosphate part of ATP is highly negatively charged, the binding free energy in the former case is much lower than that in the latter case (Fig. 2C--D). Thus, the Arg prefers binding to the triphosphate part of the ATP. We also compared the binding preference in the case of multiple Arg residues. Similarly, these Arg residues are more likely to accumulate around the triphosphate part of the ATP because the average binding free energy is much lower. Notably, due to the steric effect, the average binding free energy increases with the increase of the Arg number. Importantly, the average binding energy increases significantly, and one of the binding free energies becomes positive in the case of four Arg residues (Fig. 2C), indicating that the binding of Arg to ATP is no longer energy-favorable. Thus, the maximum binding number of Arg residues (to ATP) should be three; namely, ATP may act as a trivalent binder.

We only considered a single amino acid (i.e., Arg) in the above calculation, which may be different from the real case when the Arg is in the protein chain. To this end, we further considered a tripeptide model (i.e., Gly--Arg--Gly) and calculated the binding free energy between the tripeptide and the ATP. The use of Gly as the terminal of the tripeptide is due to the fact that Gly is the most frequently neighboring amino acid of the Arg in FUS (there are two arginine--glycine rich regions in FUS). Interestingly, Gly is a very small amino acid; thus, it could have little impact on the interaction between the Arg and the ATP. As shown in Supplementary Fig. S1, the binding free energy of the ATP--tripeptide interaction is similar to that of ATP--Arg interaction at L=1, 2, 3 (Fig. 2C), while some increase in the binding free energy is observed at L=4, probably due to the steric effect. Generally, the neighboring amino acids may not obviously affect the binding strength and the binding valance of the Arg--ATP interaction in FUS.

\textbf{Mean-field theory on the relationship between microscopic interactions and macroscopic phase behaviors.}  According to the above SQM results, the strength of ATP--Arg binding is stronger than that of Tyr--Arg binding, implying that the phase separation of FUS driven by Tyr--Arg binding could be affected by the addition of ATP. As such, to investigate the effect of ATP on the phase separation in FUS solutions, we further developed the mean-field theory by taking one-to-multiple binding between the ATP and Arg residues into account; this was inspired by the idea of dealing with specific interactions in polymer systems \cite{the1,the2,the4} (see details of the Theoretical modeling in the Materials and Methods section).

Figure 3A shows the phase diagram with critical points in the presence of ATP. Interestingly, the saturation concentration (i.e., the left part of the binodal curve separated by the critical point) of FUS shows a non-monotonic manner with the addition of ATP when ATP serves as a multivalent binder (i.e., L=2 or 3). When the amount of ATP is small, the saturation concentration decreases with the increase of ATP. However, the saturation concentration increases when the ATP concentration reaches the order of $mM$. As illustrated in Fig. 3B, ATP can bind to more than one Arg residues when the ATP concentration is low, which could be taken as an additional driving force for promoting the aggregation of proteins and causing phase separation in very dilute solutions. However, at a high ATP concentration, each ATP can only bind to one Arg because of the limited number of Arg residues, which leads to the breakage of Tyr--Arg binding and the inhibition of phase separation. This two-stage phenomenon is in good agreement with previous experiments on the effect of ATP on phase separation \cite{kang,song}: first the enhancement and the subsequent inhibition with the increase of additives. On the contrary, if ATP is just a monovalent binder (i.e., L = 1), the saturation concentration of FUS monotonically increases with the increase of ATP concentration. In other words, a higher concentration of protein solution is required for the phase separation in the presence of ATP, indicating that the ATP only inhibited phase separation. Therefore, a multivalent binder acted on by ATP is a prerequisite of the two-stage phenomenon.

 It was reported that the dissolution of protein droplets can be observed when the ATP concentration reached its physiological concentration \cite{science1}.  As shown in Fig. 3C, the protein condensates can completely disappear when the ATP concentration is about $10 mM$ for L=2 and $15 mM$ for L=3, which agrees with experimental observations ($\sim$ 8--12 mM) by Patel et al. \cite{science1}. In comparison, the  protein condensates dissolve when ATP concentration reached 0.001 mM under the condition of L=1. These results indicate that the multivalence of ATP may be an important reason why the physiological concentration of ATP is in the order of $mM$. More interestingly, with the increase in ATP, the concentration of the protein condensate $\rho_2$ first increases and then decreases in the case of L=3. This means that the addition of a small amount of ATP can not only stabilize the protein droplet, but also make the droplet denser when the ATP has a high valence (L=3). Moreover, as shown in Fig. 3D--E, both the extent of Tyr--Arg binding and the average number of bound Arg residues per ATP show a similar way of changing with the addition of ATP in the case of L=2 or 3, which is quite different from that in the case of L=1. In the latter case, one molecule of ATP always links to one Arg, which strongly decreases the Tyr--Arg binding and shows the inhibition of phase separation with ATP concentration much lower than the physiological concentration. This indicates that the changeable binding ability of ATP (to Arg) is crucial for dissolving the protein droplets at a millimolar ATP concentration.

\textbf{Coarse-grained molecular dynamics simulation on the effect of ATP on the phase separation from the dynamic view}. We also employed the coarse-grained molecular dynamics (CGMD) simulation (see details of the CG modeling in the Materials and Methods section) to investigate the effect of ATP on the phase separation. As shown in Supplementary Fig. S2A--B, the reentrant phase separation in the diluted solution is also observed once the ATP can bind to more than one Arg residues. Moreover, a large number of ATP molecules can indeed dissolve the condensates at a high concentration of protein solution (Supplementary Fig. S2C--D). Importantly, protein condensates are dynamic, having the characteristic of the exchange of protein molecules inside and outside the condensate. As shown in Movies S1--S3 in the Supplementary Materials, the phase separation is indeed in a dynamic equilibrium and the protein chains in the surrounding continuously exchange with protein chains in the droplet.  In more detail, there are many exchanged chains in the absence of ATP (Fig. 4A). With the addition of a small amount of ATP, since the protein condensate becomes denser, the number of chains participating in the exchange decreases (Fig. 4B). With the further addition of ATP, the protein condensate becomes looser and the molecular exchange is promoted again (Fig. 4C). The quantitative data of the total number of exchanged chains in a given time can be found in Supplementary Fig. S3. Generally, from the dynamic point of view, the influence of ATP on phase behavior is also non-monotonic.

\textbf{Predictions on phase separation induced by ATP for prion-Like proteins with asymmetric numbers of Tyr and Arg residues.} After revealing the non-monotonic effect of the ATP on the phase separation in the FUS solution, we further explored a more general case in prion-like proteins, in which the number of Tyr and Arg residues is not necessarily equal \cite{FUS4}. To denote such a difference, here for the sake of simplicity, we only considered two cases: the number of Tyr residues being much larger than ($m_1=34$, $m_2=11$) or smaller than ($m_1=11$, $m_2=34$) that of Arg residues. Due to the asymmetric number of Tyr and Arg residues, the phase separation is hardly observed in the pure protein solutions. With the addition of ATP, the phase separation could be observed as two looped phase diagrams (see Fig. 5A). However, the area of the loop is quite different; namely, the area in the case of $m_1 <m_2$ is much larger than that in the case of $m_1 > m_2$. This difference mainly originated from the ability of ATP binding to Arg; in other words, the average number of bound Arg residues per ATP in the case of $m_1<m_2$ is much larger than that in the case of $m_1>m_2$ (Fig. 5B). Furthermore, with a small amount of ATP, the Tyr-Arg binding is enhanced in the case of  $m_1<m_2$, which promotes phase separation  (see the inset of Fig. 5B). The above theoretical results are also verified by the CGMD simulations (Supplementary Fig. S4).  Generally, the phase separation could also be modulated by the ATP in the protein systems with asymmetrical numbers of Tyr and Arg residues, but the sensitivity of the phase separation (to the ATP concentration) is highly related to the number of Arg residues.

\section{Discussion}

Why the ATP concentration in cells is so high has long been a puzzling problem. When Patel et al. \cite{science1} found that ATP can work as a biological hydrotrope in 2017, researchers began to pay special attention to the effect of ATP on the dissolution of protein condensates \cite{hydrotrope2}.  Later, the dual roles of ATP on phase separation were reported \cite{plos,kang,song,dual}, which again put forward the question of how ATP works.

In this work, combining the effort of the SQM method, mean-field theory, and molecular simulation, we perform a systematic study on the modulation of FUS condensates aroused by ATP. The SQM method shows that Arg prefers binding to the triphosphate part of the ATP, with the binding free energy in the order of $k_BT$ and a maximum binding number of three. Taking these microscopic features as the input, we develop a mean-field theory to provide macroscopic phase diagrams based on the thermodynamic criterion; this demonstrates that the non-monotonic modulation largely originates from the distinct number of binding Arg residues (to ATP) at different ATP concentrations. Furthermore, we employ the CGMD simulation to demonstrate that the dynamic molecular exchange between the condensate and the surrounding also shows a non-monotonic behavior as the ATP concentration increases.

It should be mentioned that the effects of ligands on the phase separation of proteins may be also the opposite (i.e., enhancement or inhibition), which can be well described within the framework of polyphasic linkage first developed by Wyman and Gill \cite{ligand1}. Recently, Ruff et al. \cite{ligand2,ligand3} reported that the strength, valence, structure, and concentration of the ligand can generate important and subtle effects on the phase behaviors of multivalent macromolecules with sticker and spacer architectures. Notably, they were the first to show that monovalent ligands destabilize phase separation driven by homotypic interactions, and more importantly, the non-monotonic behavior of ligands as a function of the ligand concentration was also anticipated and demonstrated \cite{ligand2,ligand3}. Very recently, Dao et al. \cite{ligand4} found that the type of the polyubiquitin chain could also play distinct roles in the phase separation of UBQLN2, which further bolsters the general conceptual foundations of the findings from Ruff et al. These results share some similarity with the dual effect of ATP (on the phase separation) reported here, which provides a key conceptual lynchpin for our current work.

In general, this work well explains the non-monotonic effect of ATP on the phase separation. This sheds some lights on fine-tuning the phase behaviors of prion-like proteins with additional molecules and may have implications for devastating neurodegenerative disorders.

\section{Materials and Methods}

\textbf{The semiempirical quantum mechanical method}.
Here, the latest SQM method proposed by Grimme et al., the GFN2-xTB method \cite{sqm1}, is used to calculate the free energy of the Arg binding to the Tyr and/or the ATP. The accuracy of this SQM method in binding energy calculation is believed to be close to that of the density functional theory (DFT), but it has a much lower computing cost \cite{sqm2}. The binding free energy is calculated as $\Delta G=G_{com}-G_A-G_B$, where $G_{com}$ is the free energy of the complex, $G_A$ is the free energy of the Arg, and $G_B$ is the free energy of the Tyr or the ATP. The free energy G usually comes from three sources: the total molecular gas-phase energy, the corresponding free solvation energy, and the thermostatistical contribution to the free energy including the translation, rotation, vibration, and conformation degrees of freedom at a given temperature \cite{sqm1,sqm2}. To obtain the representative configuration, we first used the Molclus program \cite{sqm3} to search the possible binding structures in each case. Then, we used the GFN2-xTB method to make a preliminary optimization of the searched structures. Next, we divided the optimized structure into a complex and two monomers, and performed the Hessian calculation. All the calculations were performed under the implicit solvent model, namely using the generalized Born (GB) model with surface area (SA) contributions (i.e., GBSA(H2O)) \cite{sqm4}. The SQM calculations were performed using the xtb program \cite{sqm5}.

\textbf{Theoretical modeling of FUS solutions in the absence of ATP.}
In the theory, FUS proteins were modeled as linear polymers with a length of $N_1=526$. Because FUS belongs to a class of biomolecules with limited aqueous solubility \cite{pappu1}, solvents were modeled as chains with lengths of $N_2$, which is a simple way to reflect the limited solubility of biomolecules \cite{PRL1,PNAS1}. Each FUS protein contains some Tyr and Arg residues \cite{FUS4}. The number of Tyr or Arg residues is named as the valence of the multivalent interaction. According to experimental data from Wang et al. \cite{FUS4}, FUS has $m_1$ Tyr residues and $m_2$ Arg residues, with $m_1=m_2=34$. The incompressible solution consists of $n_p$ FUS proteins and $n_s$ solvent molecules. The dimensionless free energy of the reference state $\beta F_{ref}$ was written as:
\begin{eqnarray}
\beta F_{ref} &=& \frac{\phi}{N_1} \ln \phi + \frac{(1-\phi)}{N_2} \ln  (1-\phi)
\end{eqnarray}
where $\phi=\frac{n_p N_1 v}{V}$ is the volume fraction of FUS. $v$ is the volume of one residue of the protein, which is taken as the unit volume. $V=n_p N_1 v +n_s N_2 v$ is the volume of system. Thus, the reference system is always homogeneous, which is also the system without including Tyr--Arg binding. Next, we took the specific binding between Tyr and Arg residues into account, assuming that each Tyr/Arg has only one binding site. According to previous works on associative polymers \cite{the1,pappu3}, the free energy of Tyr--Arg binding can be written as:
\begin{eqnarray}
\beta F_{binding} &=& -\frac{v}{V} \ln Z_{binding}
\end{eqnarray}
where $Z_{binding}$ is a partition function and equals to
\begin{eqnarray}
Z_{binding} &=& P_{binding} W \exp(-k\beta\epsilon_1)
\end{eqnarray}
Here, $P_{binding}$ is a combinatorial factor that describes the number of ways to form $k$ Tyr--Arg bonds, which can be written as:
\begin{eqnarray}
P_{binding} &=& C^k_{n_p m_1} C^k_{n_p m_2} k!
\end{eqnarray}
The formation of each bond requires the Tyr and the Arg to be in close proximity within some volume scale $v_b$. For simplicity, we take $v_b$ to simply be the residue volume $v$. The overall probability of forming $k$ bonds is then
\begin{eqnarray}
W&=&(\frac{v}{V})^k
\end{eqnarray}  
In Eq. (3), $\beta\epsilon_1$ is the free energy associated with the binding between Tyr and Arg, which includes the affinity between Tyr and Arg residues, as well as the entropic loss during the formation of the Tyr--Arg bond \cite{the2}. More specifically, the specific chemical structures of Tyr and Arg side chains appear to be important determinants of the formation of the Tyr-Arg bond \cite{FUS4}. Such a reduction of conformational entropy is a common character for specific interactions, such as hydrogen bonds and cation-$\pi$ interactions. Substituting Eqs.(3)--(5) into Eq. (2), the dimensionless binding free energy can be written as:
\begin{eqnarray}
\beta F_{binding} &=& \frac{m_1\phi}{N_1} [p \ln p + (1-p) \ln (1-p)]  + \frac{m_2\phi}{N_1} \left(1-p\frac{m_1}{m_2}\right) \ln \left(1-p\frac{m_1}{m_2}\right) \nonumber\\ \
&-& p \frac{m_1\phi}{N_1}\left( \ln \frac{m_2\phi}{N_1 e} -\beta\epsilon_1\right)
\end{eqnarray}
where $p =\frac{k}{n_p m_1}$ is the fraction of Tyr in the formation of Tyr-Arg bonds and reflects the extent of protein network crosslinked by the Tyr--Arg bonds. Thus, the total free energy of the FUS solution is $F_{tot}= F_{ref}+F_{binding}$, which includes the two order parameters $\phi$ and $p$. According to the thermodynamic criterion, phase diagrams can be obtained from the equality of chemical potential and the osmotic pressure of two coexisting phases (see the details in the Supplementary Materials). The two coexisting phases consist of a diluted phase and a condensed phase, with distinct volume fractions of $\phi_1$ and $\phi_2$ respectively. It should be mentioned that the volume fraction of the diluted phase ($\phi_1$) can be converted to the saturation concentration ($\rho_1$) by considering that the hydration size of the amino acid is about 0.65 nm \cite{size}. This can be then mapped with the experimental saturation concentration (i.e., $5\mu M$) \cite{FUS4} to determine the fitting parameter $N_2=65$ in the theory.

\textbf{Theoretical modeling of FUS solutions in the presence of ATP.} The ATP is modeled as a small molecule with a volume of $v$ (the unit volume), which is the same as one residue of FUS. The reference system is chosen as a FUS solution with freely moving ATP molecules. The dimensionless free energy of the reference state can be written as:
\begin{eqnarray}
\beta\mathscr{F}_{ref}&=&\frac{\phi}{N_1} \ln \phi + \frac{(1-\lambda\phi-\phi)}{N_2} \ln (1-\lambda\phi-\phi) + \lambda\phi \ln \lambda\phi  \nonumber\\ \
&+&\chi\lambda\phi(1-\phi-\lambda\phi) -\lambda\phi \mu_{ATP}
\end{eqnarray}
where $\lambda=\frac{n_{ATP}}{n_p N_1}$ is the ratio of the number of ATP molecules and the number of FUS residues. $n_{ATP}$ is the number of ATP molecules in the solution. The first three terms are mixing the entropy for the three components of FUS, solvent and ATP. The fourth term reflects the effective interaction between ATP and the solvent with $\chi=-1.2$, since ATP can act as a hydration mediator due to its triphosphate with a unique hydration property \cite{song}. The last term is introduced to change the system into a semiclosed system \cite{the4}, which is more convenient for calculating the phase diagram. This means that ATP molecules in protein solutions are free to exchange with a reservoir at a fixed chemical potential ($\mu_{ATP}$) determined by ATP volume fraction ($\varphi$), i.e., $\mu_{ATP}=\ln \varphi - \frac{1}{N_2}\ln (1-\varphi)-\frac{1}{N_2}+1+\chi(1-2\varphi)$. The ATP volume fraction ($\varphi$) can be converted into the ATP concentration ($c$) by $c=\frac{\varphi}{v}$. Next, we take both the Tyr--Arg binding and ATP--Arg binding into account. The former is a one-to-one binding, and the latter may be one-to-multiple binding. $L$ is used to represent the valence of ATP.  According to previous works on one-to-multiple binding \cite{the2,the4}, the free energy of the formation of different kinds of bonds is
\begin{eqnarray}
\beta\mathscr{F}_{binding}&=&- \frac{v}{V} \ln \mathscr{Z}_{binding}
\end{eqnarray}
$\mathscr{Z}_{binding}$ is the partition function and is given by:
\begin{eqnarray}
\mathscr{Z}_{binding}&=&\mathscr{P}_{binding} \mathscr{W} \exp(-k\beta\epsilon_1-b\beta\epsilon_2)
\end{eqnarray}
Here $\mathscr{P}_{binding}$ is a combinatorial factor that describes the number of ways to form $k$ Tyr--Arg bonds and $b$ ATP--Arg bonds in the system with $n_p$ FUS proteins and $n_{ATP}$ ATP molecules. It can be written as:
\begin{eqnarray}
\mathscr{P}_{binding}&=& C^k_{n_p m_1}C^b_{n_{ATP} L} C^{k+b}_{n_p m_2}(k+b)!
\end{eqnarray}
$\mathscr{W}$ in Eq. (9) is a probability of finding $(k+b)$ bonds in the vicinity of each other, which is
\begin{eqnarray}
\mathscr{W} &=& (\frac{v}{V})^{(k+b)}
\end{eqnarray} 
In Eq. (9), $\beta\epsilon_2$ is the binding free energy aroused by the formation of one ATP--Arg bond. Substituting Eqs. (9)--(11) into Eq. (8), the associated dimensionless free energy is given by:
\begin{eqnarray}
\beta\mathscr{F}_{binding}&=&\frac{m_1\phi}{N_1} [p\ln p +(1-p)\ln (1-p)] + L\lambda\phi [q\ln q +(1-q)\ln (1-q)] \nonumber\\ \
&+& \frac{m_2 \phi}{N_1} \left( 1-\frac{p m_1}{m_2}-\frac{q L \lambda N_1}{m_2} \right) \ln \left(1- \frac{p m_1}{m_2} - \frac{q L \lambda N_1}{m_2}\right) \nonumber\\ \
&-&p \phi \frac{m_1}{N_1} \left(\ln \frac{m_2 \phi}{N_1 e} -\beta\epsilon_1 \right) -qL\lambda \phi \left(\ln \frac{m_2 \phi}{N_1 e}-\beta \epsilon_2 \right)
\end{eqnarray}
where $q=\frac{b}{L \lambda n_p N_1}$ is the fraction of ATP participating in ATP--Arg binding. Therefore, the total free energy becomes $\mathscr{F}_{tot}=\mathscr{F}_{ref}+\mathscr{F}_{binding}$. It can be seen that $p$ and $q$ are coupled in Eq. (12), indicating the competition between ATP--FUS and FUS--FUS interactions. After minimizing $\mathscr{F}_{tot}$ with respect to $p$, $q$ and $\mu$, the phase diagram can be obtained according to the thermodynamic criteria of the same chemical potential and osmotic pressure of two coexisting phases (see the details in the Supplementary Materials).

\textbf{The coarse-grained molecular dynamics simulation.}
A coarse-grained molecular dynamics simulation was used to investigate the effect of ATP on the phase separation of the protein solution. The protein was modeled as a chain with connective beads in the implicit solvent \cite{wingreen1}. These beads were connected by the harmonic spring with the potential $U_b(r)=k(r-r_b)^2$, where $r_b=4.5 nm$ is the mean bond length and $k=20\epsilon_0/r_b^2$ is the bond stiffness, where $\epsilon_0$ is the energy unit. There were three types of beads in the protein chain, namely the Tyr beads, the Arg beads, and the non-interacting beads. Here, the percentage of the Tyr or Arg beads was 0.06, similar to that in the mean-field theory. To reflect the effective binding between the Tyr bead and the Arg bead, the attractive potential  $U_a(r)=U_0(1+\cos \frac{\pi r}{r_0})$ for $r < r_0$ was used. We chose $r_0=2.0 nm$ and $U_0=-20\epsilon_0$ \cite{wingreen1}. Moreover, a softened and truncated Lennard-Jones potential was adopted to avoid the aggregation of the same type of beads, which was given by: $U_r(r)=4\epsilon\lambda\{[(1-\lambda)^2+(\frac{r}{\sigma})^6]^{-2}-[(1-\lambda)^2+(\frac{r}{\sigma})^6]^{-1}\}$ when $r<r_c$. $\epsilon=0.621\epsilon_0$ and $\epsilon=0.15\epsilon_0$ were chosen for the Tyr bead and the Arg bead, respectively. Other parameters are fixed as $\lambda=0.68$, $\sigma=3.5 nm$, and $r_c=5 nm$, which are similar to those used in the previous article \cite{wingreen1}. The ATP was modeled as one bead in the simulations. The effective binding between the ATP bead and Arg bead was also modeled as the attraction potential  $U_a(r)=U_0(1+\cos \frac{\pi r}{r_0})$. Here, $r_0=2.0 nm$ and $U_0=-75\epsilon_0$ were chosen because the binding free energy of ATP--Arg was nearly four times as strong as that of Tyr--Arg according to the SQM result.

The coarse-grained simulations were performed at the NVT ensemble with the Langevin thermostat at room temperature. The size of the simulation box was $50nm\times 50nm \times 250nm$ and the periodic boundary conditions were adopted in all three directions. To promote the system to equilibrium, the slab method was adopted in the simulation \cite{Mittal2}. The time step in the equilibrium simulation was $dt=0.5 ns$. The data were collected every 50 $\mu s$, with a total time of about 50 ms. All the simulations were performed using the LAMMPS software package (29 Oct 2020) \cite{lammps}.


\section{Acknowledgements}
We are grateful to the High Performance Computing Center (HPCC) of Nanjing University for the numerical calculations in this paper on its blade cluster system.
\textbf{Funding:} This work is supported by the National Natural Science Foundation of China under Grants 11774146, 11874045, 11974175, and 12174184. 

\textbf{Author contributions:} C.-L. R., H.-M. D. and Y.-Q. M. designed and supervised research; C.-L. R. and P. Z. developed mean-field theory; Y. S. performed the coarse-grained molecular dynamics simulation; H.-M. D. performed the semi-empirical quantum mechanical method; C.-L. R., Y. S., H.-M. D. and Y.-Q. M. wrote the paper.

\textbf{Competing Interests:} The authors declare no competing interests.

\textbf{Data and materials availability:} The authors declare that the data supporting the findings of this study are available in the article and Supplementary Materials. Softwares including xtb 6.3.3 (https://github.com/grimme-lab/xtb), Molclus 1.9.9  \\
(http://www.Keinsci.Com/research/molclus.html), and LAMMPS (29 Oct 2020) \\
(https://www.lammps.org/) used in this work can be downloaded free of charge from their official websites.

\section{Supplementary Materials}

Supplementary Text;
Movies S1 to S3;
Figs. S1 to S4.  \\
Supplementary material for this article is available at https://science.org/doi/XXXXXX.

\section{Figures}
\newpage

\begin{figure*}
  \begin{center}
    \begin{tabular}{lr}
      \resizebox{150mm}{!}{\includegraphics{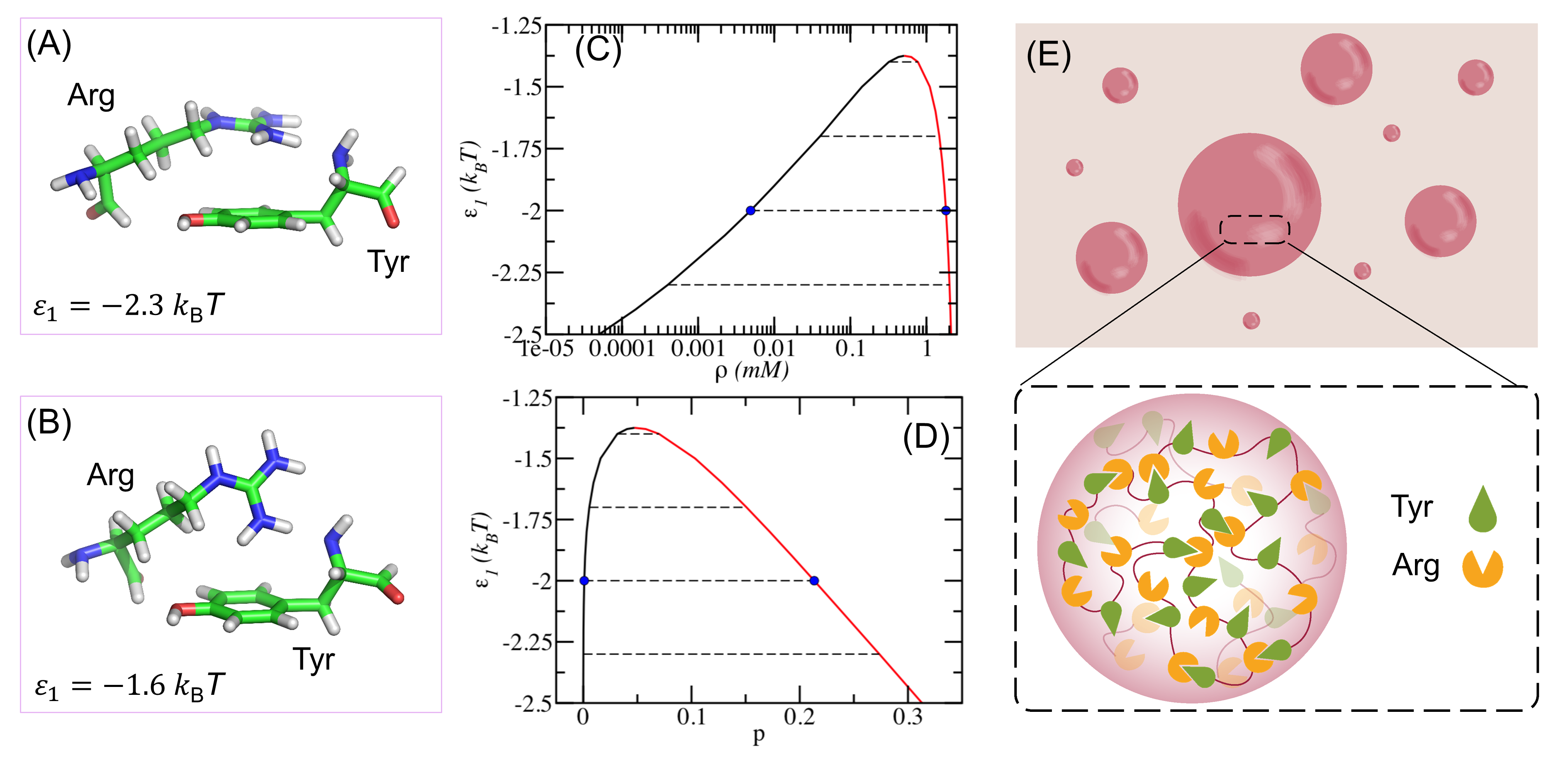}}
       \end{tabular}
  \end{center}
  \caption{Tyr--Arg binding resulting in the phase separation in FUS solution. Snapshots for the two typical binding modes between one Tyr and one Arg obtained by the SQM method (A)--(B), where $\epsilon_1$ denotes the binding free energy in each case. Phase diagram and the fraction of Tyr-Arg binding calculated from the mean-field theory (C)--(D). The black curve represents the diluted phase, and the red one represents the dense phase. Dash lines are tie-lines relating two coexisting phases. The blue dots at $\epsilon_1=-2 k_BT$ are taken as the average binding free energy between Tyr and Arg for the following theoretical calculations. (E) Schematic representation of the formation of FUS protein droplets caused by Tyr--Arg binding.}.
\end{figure*}

\begin{figure*}
  \begin{center}
    \begin{tabular}{lr}
      \resizebox{130mm}{!}{\includegraphics{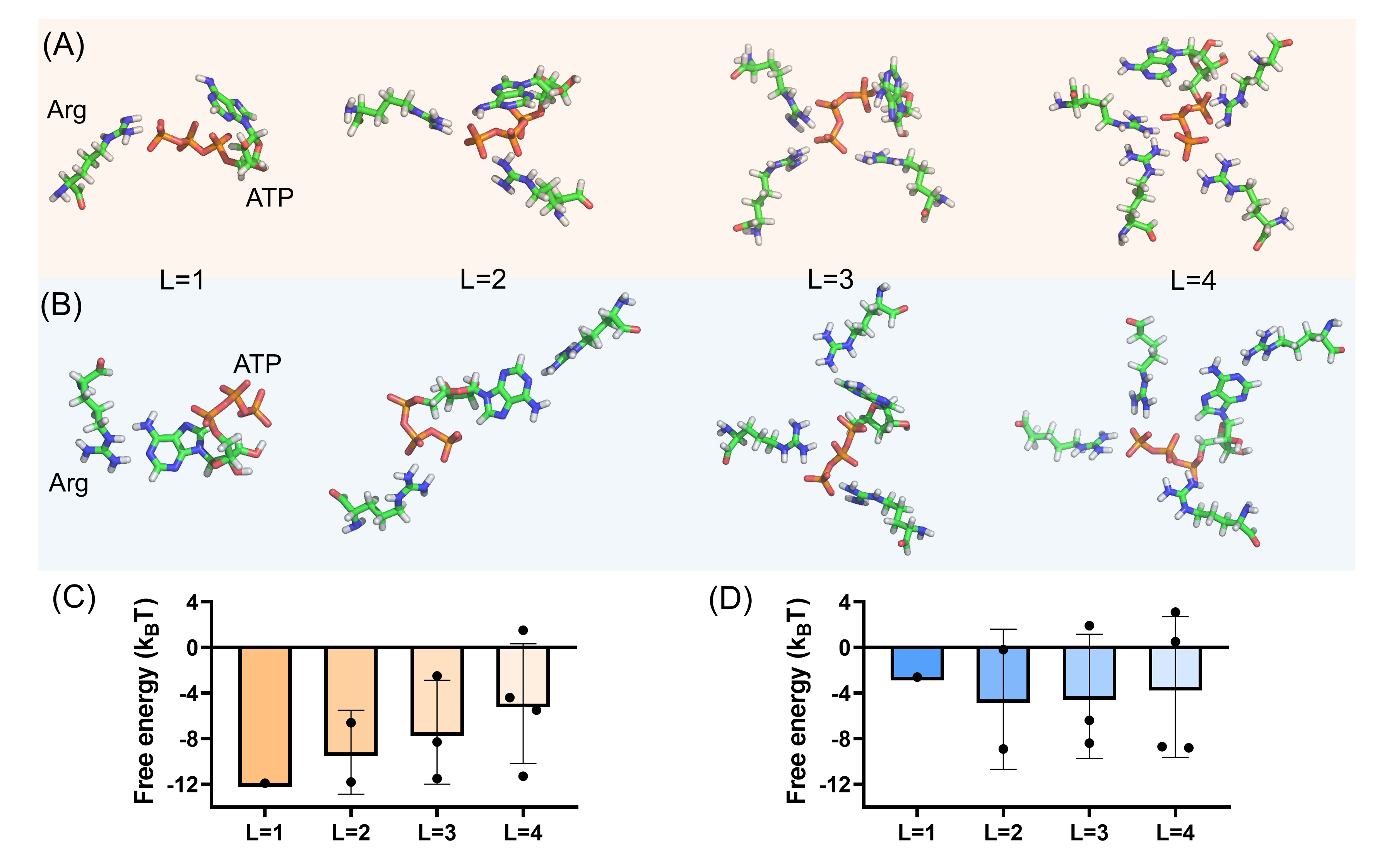}}
       \end{tabular}
  \end{center}
  \caption{The SQM result for the ATP--Arg binding. Snapshots for the typical binding modes between one ATP and one/several Arg residues: (A) the Arg residue(s) all bind to the triphosphate part of the ATP; (B) one of the Arg residue(s) binds to the adenine part of the ATP.  The number of Arg residues is denoted as the valence of ATP, represented by $L$. (C)--(D) shows the corresponding binding free energy of the ATP--Arg interaction as a function of $L$ in the case of (A) and (B), respectively.}
\end{figure*}

\begin{figure*}
  \begin{center}
    \begin{tabular}{lr}
      \resizebox{160mm}{!}{\includegraphics{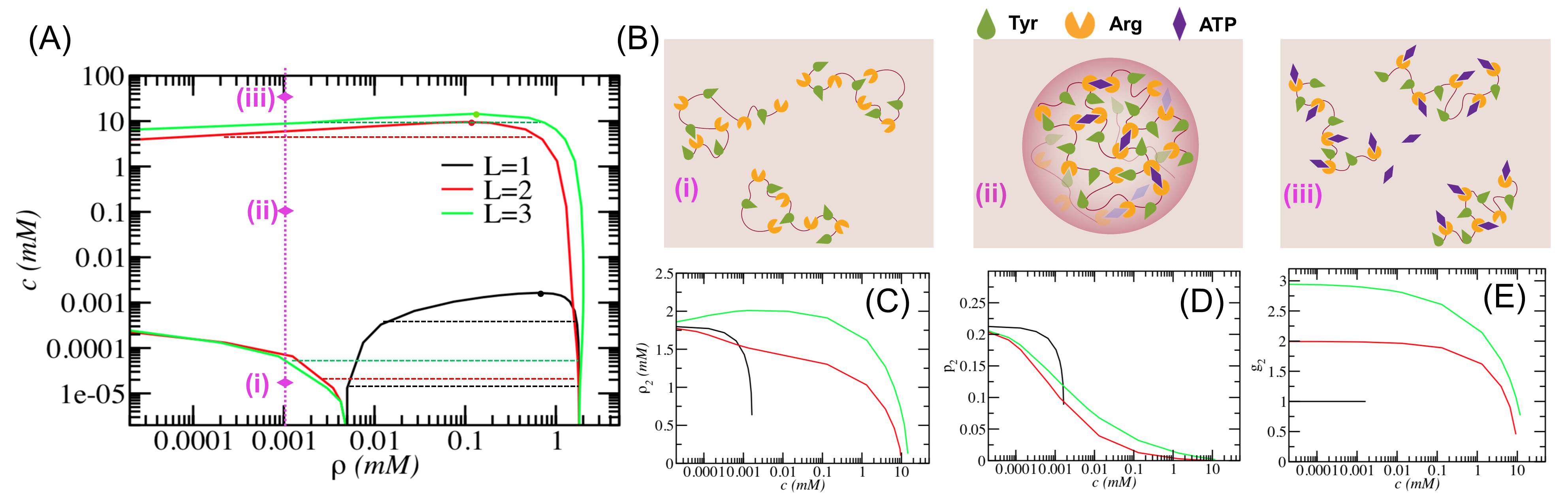}}
      \end{tabular}
  \end{center}
  \caption{The reentrant phase separation and protein dissolution in mean-field theory. (A) Phase diagram of FUS solutions as a function of ATP concentration ($c$) for three different ATP valences $L$. The parameter $\beta\epsilon_2$ at L=1, 2, 3 is set based on the average binding free energy in Fig. 2C. Dashed lines are tie-lines relating two coexisting phases. Dots are critical points of each cases. The dot line with diamond symbols (i)--(iii) indicates the reentrant phase transition with the increase of ATP concentration. (B) Schematic representations for (i)--(iii) the reentrant phase transition occurring in dilute solutions for L=2. (C) The concentration of protein condensates ($\rho_2$), (D) the extent of Tyr--Arg binding ($p_2$), and (E) the average number of bound Arg residues per ATP ($g_2$)  in protein condensates as functions of the ATP concentration.}
\end{figure*}

 \begin{figure*}
  \begin{center}
    \begin{tabular}{lr}
      \resizebox{160mm}{!}{\includegraphics{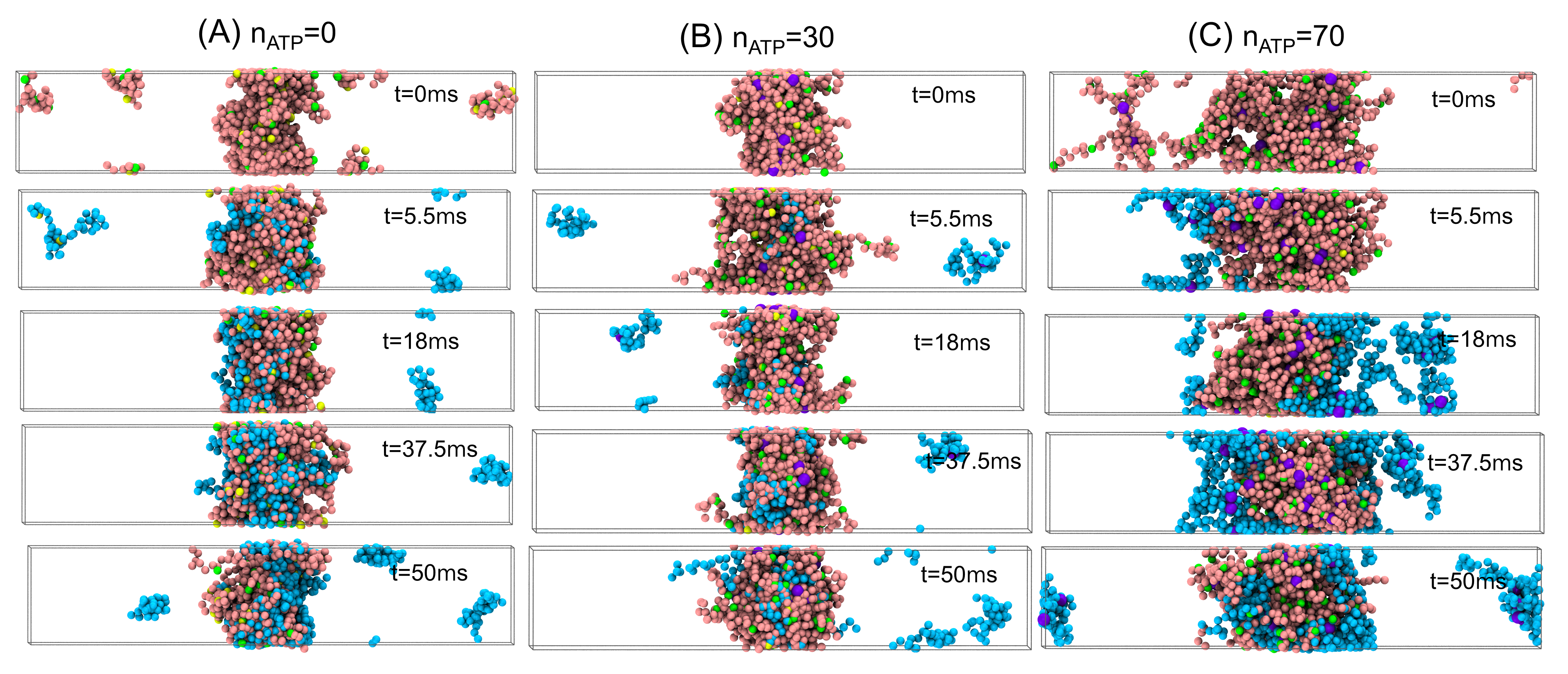}}
    \end{tabular}
  \end{center}
  \caption{Molecular exchange of protein condensates. Time sequence of snapshots illustrating the  exchange of protein chains inside and outside the protein condensates in CGMD simulations: (A) $n_{ATP}=0$, (B) $n_{ATP}=30$, (C) $n_{ATP}=70$. The protein chains participating in the molecular exchange are shown in cyan, and others are described as pink chains. The yellow and green beads represent Arg and Tyr residues respectively, the purple beads stands for the ATP molecules. The total number of protein chains is $n_{pro}=50$ in the simulations.}
\end{figure*}

\begin{figure*}
  \begin{center}
    \begin{tabular}{lr}
     \resizebox{140mm}{!}{\includegraphics{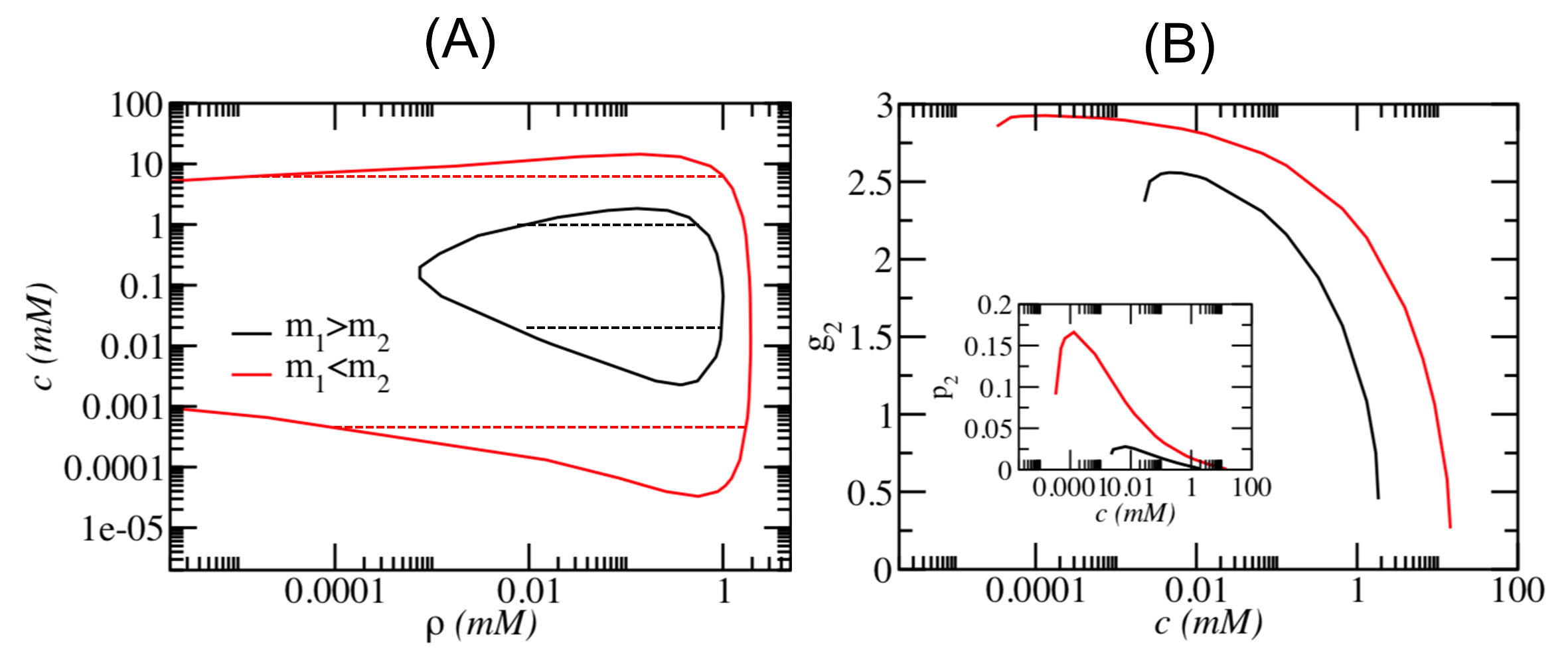}}
     \end{tabular}
  \end{center}
  \caption{The effect of ATP on other prion-like proteins. (A) Phase diagram of prion-like proteins with asymmetric numbers of Tyr and Arg residues ($m_1 > m_2$ and $m_1 < m_2$) with the addition of ATP for L=3. (B) The average number of bound Arg residues per ATP ($g_2$) and the extent of Tyr--Arg binding ($p_2$) (inset) in protein condensates as functions of ATP concentration.}
\end{figure*}

\end{document}